\documentclass{iopconfser}
\usepackage[acronym]{glossaries}
\usepackage{graphicx} 
\usepackage{dcolumn} 
\usepackage{colordvi}
\usepackage{color}
\usepackage{epstopdf}
\usepackage{amssymb}
\usepackage{url}
\graphicspath{{ps}}
\usepackage{hyperref}
\usepackage{tabularx}
\usepackage{xpatch}
\usepackage{hepparticles}
\usepackage{hepnames}
\usepackage{lineno}
\usepackage[nameinlink, capitalise]{cleveref}
\usepackage{siunitx}
\usepackage{tikz}
\usepackage{xargs} 
\usepackage{placeins}
\usepackage{amsmath}
\usepackage{slashed}

\usepackage{float}
\usepackage[
    style=numeric,
    maxnames=3,
    minnames=1,
    maxcitenames=2,
    mincitenames=1
]{biblatex}
\begin{document}


\def\belletwo {Belle~II }

\let\emi\en
\def\electron   {\ensuremath{e}\xspace}
\def\en         {\ensuremath{e^-}\xspace}   
\def\ep         {\ensuremath{e^+}\xspace}
\def\epm        {\ensuremath{e^\pm}\xspace} 
\def\epem       {\ensuremath{e^+e^-}\xspace}
\def\ee         {\ensuremath{e^-e^-}\xspace}

\def\mmu        {\ensuremath{\mu}\xspace}
\def\mup        {\ensuremath{\mu^+}\xspace}
\def\mun        {\ensuremath{\mu^-}\xspace} 
\def\mumu       {\ensuremath{\mu^+\mu^-}\xspace}
\def\mtau       {\ensuremath{\tau}\xspace}
\def\mupm       {\ensuremath{\mu^\pm}\xspace}

\def\taup       {\ensuremath{\tau^+}\xspace}
\def\taum       {\ensuremath{\tau^-}\xspace}
\def\tautau     {\ensuremath{\tau^+\tau^-}\xspace}

\def\ellm       {\ensuremath{\ell^-}\xspace}
\def\ellp       {\ensuremath{\ell^+}\xspace}
\def\ellell     {\ensuremath{\ell^+ \ell^-}\xspace}

\def\nub        {\ensuremath{\overline{\nu}}\xspace}
\def\nunub      {\ensuremath{\nu{\overline{\nu}}}\xspace}
\def\nub        {\ensuremath{\overline{\nu}}\xspace}
\def\nunub      {\ensuremath{\nu{\overline{\nu}}}\xspace}
\def\nue        {\ensuremath{\nu_e}\xspace}
\def\nueb       {\ensuremath{\nub_e}\xspace}
\def\nuenueb    {\ensuremath{\nue\nueb}\xspace}
\def\num        {\ensuremath{\nu_\mu}\xspace}
\def\numb       {\ensuremath{\nub_\mu}\xspace}
\def\numnumb    {\ensuremath{\num\numb}\xspace}
\def\nut        {\ensuremath{\nu_\tau}\xspace}
\def\nutb       {\ensuremath{\nub_\tau}\xspace}
\def\nutnutb    {\ensuremath{\nut\nutb}\xspace}
\def\nul        {\ensuremath{\nu_\ell}\xspace}
\def\nulb       {\ensuremath{\nub_\ell}\xspace}
\def\nulnulb    {\ensuremath{\nul\nulb}\xspace}


\def\g     {\ensuremath{\gamma}\xspace}
\def\gaga  {\ensuremath{\gamma\gamma}\xspace}  
\def\ggstar{\ensuremath{\gamma\gamma^*}\xspace}

\def\ega    {\ensuremath{e\gamma}\xspace}
\def\game   {\ensuremath{\gamma e^-}\xspace}
\def\epemg  {\ensuremath{e^+e^-\gamma}\xspace}


\def\H      {\ensuremath{H^0}\xspace}
\def\Hp     {\ensuremath{H^+}\xspace}
\def\Hm     {\ensuremath{H^-}\xspace}
\def\Hpm    {\ensuremath{H^\pm}\xspace}
\def\W      {\ensuremath{W}\xspace}
\def\Wp     {\ensuremath{W^+}\xspace}
\def\Wm     {\ensuremath{W^-}\xspace}
\def\Wpm    {\ensuremath{W^\pm}\xspace}
\def\Z      {\ensuremath{Z^0}\xspace}


\def\q     {\ensuremath{q}\xspace}
\def\qbar  {\ensuremath{\overline q}\xspace}
\def\qqbar {\ensuremath{q\overline q}\xspace}
\def\u     {\ensuremath{u}\xspace}
\def\ubar  {\ensuremath{\overline u}\xspace}
\def\uubar {\ensuremath{u\overline u}\xspace}
\def\d     {\ensuremath{d}\xspace}
\def\dbar  {\ensuremath{\overline d}\xspace}
\def\ddbar {\ensuremath{d\overline d}\xspace}
\def\s     {\ensuremath{s}\xspace}
\def\sbar  {\ensuremath{\overline s}\xspace}
\def\ssbar {\ensuremath{s\overline s}\xspace}
\def\c     {\ensuremath{c}\xspace}
\def\cbar  {\ensuremath{\overline c}\xspace}
\def\ccbar {\ensuremath{c\overline c}\xspace}
\def\b     {\ensuremath{b}\xspace}
\def\bbar  {\ensuremath{\overline b}\xspace}
\def\bbbar {\ensuremath{b\overline b}\xspace}
\def\t     {\ensuremath{t}\xspace}
\def\tbar  {\ensuremath{\overline t}\xspace}
\def\tbar  {\ensuremath{\overline t}\xspace}
\def\ttbar {\ensuremath{t\overline t}\xspace}


\def\piz   {\ensuremath{\pi^0}\xspace}
\def\pizs  {\ensuremath{\pi^0\mbox\,\rm{s}}\xspace}
\def\ppz   {\ensuremath{\pi^0\pi^0}\xspace}
\def\pip   {\ensuremath{\pi^+}\xspace}
\def\pim   {\ensuremath{\pi^-}\xspace}
\def\pipi  {\ensuremath{\pi^+\pi^-}\xspace}
\def\pipm  {\ensuremath{\pi^\pm}\xspace}
\def\pimp  {\ensuremath{\pi^\mp}\xspace}

\def\kaon  {\ensuremath{K}\xspace}
\def\Kbar  {\kern 0.2em\overline{\kern -0.2em K}{}\xspace}
\def\Kb    {\ensuremath{\Kbar}\xspace}
\def\Kz    {\ensuremath{K^0}\xspace}
\def\Kzb   {\ensuremath{\Kbar^0}\xspace}
\def\KzKzb {\ensuremath{\Kz \kern -0.16em \Kzb}\xspace}
\def\Kp    {\ensuremath{K^+}\xspace}
\def\Km    {\ensuremath{K^-}\xspace}
\def\Kpm   {\ensuremath{K^\pm}\xspace}
\def\Kmp   {\ensuremath{K^\mp}\xspace}
\def\KpKm  {\ensuremath{\Kp \kern -0.16em \Km}\xspace}
\def\KS    {\ensuremath{K^0_{\scriptscriptstyle S}}\xspace} 
\def\KL    {\ensuremath{K^0_{\scriptscriptstyle L}}\xspace} 
\def\Kstarz  {\ensuremath{K^{*0}}\xspace}
\def\Kstarzb {\ensuremath{\Kbar^{*0}}\xspace}
\def\Kstar   {\ensuremath{K^*}\xspace}
\def\Kstarb  {\ensuremath{\Kbar^*}\xspace}
\def\Kstarp  {\ensuremath{K^{*+}}\xspace}
\def\Kstarm  {\ensuremath{K^{*-}}\xspace}
\def\Kstarpm {\ensuremath{K^{*\pm}}\xspace}
\def\Kstarmp {\ensuremath{K^{*\mp}}\xspace}

\newcommand{\etapr}{\ensuremath{\eta^{\prime}}\xspace}


\def\Dbar    {\kern 0.2em\overline{\kern -0.2em D}{}\xspace}
\def\Db      {\ensuremath{\Dbar}\xspace}
\def\Dz      {\ensuremath{D^0}\xspace}
\def\Dzb     {\ensuremath{\Dbar^0}\xspace}
\def\DzDzb   {\ensuremath{\Dz {\kern -0.16em \Dzb}}\xspace}
\def\Dp      {\ensuremath{D^+}\xspace}
\def\Dm      {\ensuremath{D^-}\xspace}
\def\Dpm     {\ensuremath{D^\pm}\xspace}
\def\Dmp     {\ensuremath{D^\mp}\xspace}
\def\DpDm    {\ensuremath{\Dp {\kern -0.16em \Dm}}\xspace}
\def\Dstar   {\ensuremath{D^*}\xspace}
\def\Dstarb  {\ensuremath{\Dbar^*}\xspace}
\def\Dstarz  {\ensuremath{D^{*0}}\xspace}
\def\Dstarzb {\ensuremath{\Dbar^{*0}}\xspace}
\def\Dstarp  {\ensuremath{D^{*+}}\xspace}
\def\Dstarm  {\ensuremath{D^{*-}}\xspace}
\def\Dstarpm {\ensuremath{D^{*\pm}}\xspace}
\def\Dstarmp {\ensuremath{D^{*\mp}}\xspace}
\def\Ds      {\ensuremath{D^+_s}\xspace}
\def\Dsb     {\ensuremath{\Dbar^+_s}\xspace}
\def\Dss     {\ensuremath{D^{*+}_s}\xspace}

\def\B       {\ensuremath{B}\xspace}
\def\Bbar    {\kern 0.18em\overline{\kern -0.18em B}{}\xspace}
\def\Bb      {\ensuremath{\Bbar}\xspace}
\def\BB      {\ensuremath{B\Bbar}\xspace} 
\def\Bz      {\ensuremath{B^0}\xspace}
\def\Bzb     {\ensuremath{\Bbar^0}\xspace}
\def\BzBzb   {\ensuremath{\Bz {\kern -0.16em \Bzb}}\xspace}
\def\Bu      {\ensuremath{B^+}\xspace}
\def\Bub     {\ensuremath{B^-}\xspace}
\def\Bp      {\ensuremath{\Bu}\xspace}
\def\Bm      {\ensuremath{\Bub}\xspace}
\def\Bpm     {\ensuremath{B^\pm}\xspace}
\def\Bmp     {\ensuremath{B^\mp}\xspace}
\def\BpBm    {\ensuremath{\Bu {\kern -0.16em \Bub}}\xspace}
\def\Bs      {\ensuremath{B_s}\xspace}
\def\Bsb     {\ensuremath{\Bbar_s}\xspace}


\def\jpsi     {\ensuremath{{J\mskip -3mu/\mskip -2mu\psi\mskip 2mu}}\xspace}
\def\psitwos  {\ensuremath{\psi{(2S)}}\xspace}
\def\psiprpr  {\ensuremath{\psi(3770)}\xspace}
\def\etac     {\ensuremath{\eta_c}\xspace}
\def\chiczero {\ensuremath{\chi_{c0}}\xspace}
\def\chicone  {\ensuremath{\chi_{c1}}\xspace}
\def\chictwo  {\ensuremath{\chi_{c2}}\xspace}
\mathchardef\Upsilon="7107
\def\Y#1S{\ensuremath{\Upsilon{(#1S)}}\xspace}
\def\OneS  {\Y1S}
\def\TwoS  {\Y2S}
\def\ThreeS{\Y3S}
\def\FourS {\Y4S}
\def\FiveS {\Y5S}
\def\SixS  {\Y6S}

\def\chic#1{\ensuremath{\chi_{c#1}}\xspace}


\def\proton      {\ensuremath{p}\xspace}
\def\antiproton  {\ensuremath{\overline p}\xspace}
\def\neutron     {\ensuremath{n}\xspace}
\def\antineutron {\ensuremath{\overline n}\xspace}

\mathchardef\Deltares="7101
\mathchardef\Xi="7104
\mathchardef\Lambda="7103
\mathchardef\Sigma="7106
\mathchardef\Omega="710A

\def\Deltabar{\kern 0.25em\overline{\kern -0.25em \Deltares}{}\xspace}
\def\Lbar{\kern 0.2em\overline{\kern -0.2em\Lambda\kern 0.05em}\kern-0.05em{}\xspace}
\def\Sigbar{\kern 0.2em\overline{\kern -0.2em \Sigma}{}\xspace}
\def\Xibar{\kern 0.2em\overline{\kern -0.2em \Xi}{}\xspace}
\def\Obar{\kern 0.2em\overline{\kern -0.2em \Omega}{}\xspace}
\def\Nbar{\kern 0.2em\overline{\kern -0.2em N}{}\xspace}
\def\Xb{\kern 0.2em\overline{\kern -0.2em X}{}\xspace}

\def\X {\ensuremath{X}\xspace}


\newcommand{\tev}{\ensuremath{\mathrm{\,Te\kern -0.1em V}}\xspace}
\newcommand{\gev}{\ensuremath{\mathrm{\,Ge\kern -0.1em V}}\xspace}
\newcommand{\mev}{\ensuremath{\mathrm{\,Me\kern -0.1em V}}\xspace}
\newcommand{\kev}{\ensuremath{\mathrm{\,ke\kern -0.1em V}}\xspace}
\newcommand{\ev}{\ensuremath{\mathrm{\,e\kern -0.1em V}}\xspace}
\newcommand{\gevc}{\ensuremath{{\mathrm{\,Ge\kern -0.1em V\!/}c}}\xspace}
\newcommand{\mevc}{\ensuremath{{\mathrm{\,Me\kern -0.1em V\!/}c}}\xspace}
\newcommand{\gevcc}{\ensuremath{{\mathrm{\,Ge\kern -0.1em V\!/}c^2}}\xspace}
\newcommand{\mevcc}{\ensuremath{{\mathrm{\,Me\kern -0.1em V\!/}c^2}}\xspace}


\def\syin {\ensuremath{^{\prime\prime}}\xspace}
\def\inch   {\ensuremath{\rm \,in}\xspace} 
\def\ft   {\ensuremath{\rm \,ft}\xspace}
\def\km   {\ensuremath{{\rm \,km}}\xspace}
\def\m    {\ensuremath{{\rm \,m}}\xspace}
\def\cm   {\ensuremath{{\rm \,cm}}\xspace}
\def\cma  {\ensuremath{{\rm \,cm}^2}\xspace}
\def\mm   {\ensuremath{{\rm \,mm}}\xspace}
\def\mma  {\ensuremath{{\rm \,mm}^2}\xspace}
\def\mum  {\ensuremath{{\,\mu\rm m}}\xspace}
\def\muma       {\ensuremath{{\,\mu\rm m}^2}\xspace}
\def\nm   {\ensuremath{{\rm \,nm}}\xspace}
\def\fm   {\ensuremath{{\rm \,fm}}\xspace}
\def\nm         {\ensuremath{{\rm \,nm}}\xspace}   
\def\barn{\ensuremath{{\rm \,b}}\xspace}
\def\barnhyph{\ensuremath{{\rm -b}}\xspace}
\def\mbarn{\ensuremath{{\rm \,mb}}\xspace}
\def\mbarnhyph{\ensuremath{{\rm -mb}}\xspace}
\def\nb         {\ensuremath{{\rm \,nb}}\xspace}
\def\invnb {\ensuremath{\mbox{\,nb}^{-1}}\xspace}
\def\pb {\ensuremath{{\rm \,pb}}\xspace}
\def\invpb {\ensuremath{\mbox{\,pb}^{-1}}\xspace}
\def\fb   {\ensuremath{\mbox{\,fb}}\xspace}
\def\invfb   {\ensuremath{\mbox{\,fb}^{-1}}\xspace}
\def\invab   {\ensuremath{\mbox{\,ab}^{-1}}\xspace}


\def\mus  {\ensuremath{\rm \,\mus}\xspace}
\def\ns   {\ensuremath{\rm \,ns}\xspace}
\def\ps   {\ensuremath{\rm \,ps}\xspace}
\def\fs   {\ensuremath{\rm \,fs}\xspace}
\def\gm   {\ensuremath{\rm \,g}\xspace}

\def\sec{\ensuremath{\rm {\,s}}\xspace}       
\def\ms         {\ensuremath{{\rm \,ms}}\xspace}     
\def\mus        {\ensuremath{\,\mu{\rm s}}\xspace}    
\def\ns         {\ensuremath{{\rm \,ns}}\xspace}      
\def\ps         {\ensuremath{{\rm \,ps}}\xspace}  


\def\Xrad {\ensuremath{X_0}\xspace}
\def\NIL{\ensuremath{\lambda_{int}}\xspace}
\let\dgr\degrees

\def\cms         {\ensuremath{{\rm \,cm}^{-2} {\rm s}^{-1}}\xspace}

\def\mic  {\ensuremath{\,\mu{\rm C}}\xspace}
\def\krad {\ensuremath{\rm \,krad}\xspace}
\def\cmc  {\ensuremath{{\rm \,cm}^3}\xspace}
\def\yr   {\ensuremath{\rm \,yr}\xspace}
\def\hr   {\ensuremath{\rm \,hr}\xspace}
\def\degc {\ensuremath{^\circ}{C}\xspace}
\def\degk {\ensuremath {\rm K}\xspace}
\def\degrees{\ensuremath{^{\circ}}\xspace}
\def\mrad{\ensuremath{\rm \,mrad}\xspace}               
\def\rad{\ensuremath{\rm \,rad}\xspace}
\def\mradhyph{\ensuremath{\rm -mr}\xspace}
\def\sx    {\ensuremath{\sigma_x}\xspace}    
\def\sy    {\ensuremath{\sigma_y}\xspace}   
\def\sz    {\ensuremath{\sigma_z}\xspace}

\def\order{{\ensuremath{\cal O}}\xspace}
\def\L{{\ensuremath{\cal L}}\xspace}
\def\calL{{\ensuremath{\cal L}}\xspace}
\def\calS{{\ensuremath{\cal S}}\xspace}
\def\calA{{\ensuremath{\cal A}}\xspace}
\def\calD{{\ensuremath{\cal D}}\xspace}
\def\calR{{\ensuremath{\cal R}}\xspace}

\def\ra                 {\ensuremath{\rightarrow}\xspace}
\def\to                 {\ensuremath{\rightarrow}\xspace}

\newcommand{\stat}{\ensuremath{\mathrm{(stat)}}\xspace}
\newcommand{\syst}{\ensuremath{\mathrm{(syst)}}\xspace}

\newcommand{\inverse}{\ensuremath{^{-1}}\xspace}
\newcommand{\dedx}{\ensuremath{\mathrm{d}\hspace{-0.1em}E/\mathrm{d}x}\xspace}
\newcommand{\chisq}{\ensuremath{\chi^2}\xspace}
\newcommand{\delm}{\ensuremath{m_{\dstr}-m_{\dz}}\xspace}
\newcommand{\lum} {\ensuremath{\mathcal{L}}\xspace}

\def\gsim{{~\raise.15em\hbox{$>$}\kern-.85em
          \lower.35em\hbox{$\sim$}~}\xspace}
\def\lsim{{~\raise.15em\hbox{$<$}\kern-.85em
          \lower.35em\hbox{$\sim$}~}\xspace}

\def\qsq                {\ensuremath{q^2}\xspace}

\newcommand{\madgraph}{\texttt{MadGraph5\_aMC@NLO}}
\newcommand{\evtgen}{\texttt{EvtGen}}
\newcommand{\evtgendecays}{\texttt{EvtGenDecays}}

\newcommand{\basftwolong}{\texttt{Belle~II Analysis Software Framework}}
\newcommand{\basftwoshort}{\texttt{basf2} }

\newcommand{\treefit}{\texttt{treeFit}}

\newcommand{\zfit}{\texttt{zfit}}

\newcommand{\targetlumi}{$\int \mathcal{L}\mathrm{d}t = \SI{364}{fb^{-1}}$} 

\newcommand{\analysisrelease}{\texttt{light-2401-ocicat}}

\newcommand{\pvalue}{$p$-value}

\newcommand{\aafh}{\texttt{AAFH}}
\newcommand{\kkmc}{\texttt{KKMC}}
\newcommand{\babayaga}{\texttt{BabaYaga}}
\newcommand{\tauola}{\texttt{Tauola}}
\newcommand{\pythia}{\texttt{Pythia8}}
\newcommand{\treps}{\texttt{TREPS}}
\newcommand{\koralw}{\texttt{KoralW}}
\newcommand{\phokhara}{\texttt{PHOKHARA}}

\newcommand{\gfour}{\texttt{Geant4}}
\newcommand{\mcparticle}{\texttt{MCParticle}}
\newcommand{\mcparticles}{\texttt{MCParticles}}
\newcommand{\mcdeposition}{\texttt{MCDeposition}}
\newcommand{\mcdepositions}{\texttt{MCDepositions}}
\newcommand{\gfourtrack}{\texttt{Geant4Track}}

    \newcommand{\chitwo}{\HepParticle{\chi}{2}{}}
\newcommand{\chione}{\HepParticle{\chi}{1}{}}
\newcommand{\AChiOne}{\HepAntiParticle{\chi}{1}{}}
\newcommand{\AChiTwo}{\HepAntiParticle{\chi}{2}{}}

\newcommand{\PhiPlus}{\HepParticle{\phi}{}{+}}
\newcommand{\PhiMinus}{\HepParticle{\phi}{}{-}}

\newcommand{\hprime}{\HepParticle{h}{}{\prime}}

\newcommand{\aprime}{\HepParticle{A}{}{\prime}}

\newcommand{\eemumubkg}{\HepProcess{\Ppositron \Pelectron \to \Ppositron \Pelectron \APmuon \Pmuon}}
\newcommand{\eeeebkg}{\HepProcess{\Ppositron \Pelectron \to \Ppositron \Pelectron \Ppositron \Pelectron}}

\newcommand{\llxxbkg}{\HepProcess{\Ppositron \Pelectron \to \Pleptonplus \Pleptonminus XX}}
\newcommand{\eetautaubkg}{\HepProcess{\Ppositron \Pelectron \to \Ppositron \Pelectron \APtauon \Ptauon}}
\newcommand{\eepipibkg}{\HepProcess{\Ppositron \Pelectron \to \Ppositron \Pelectron \Ppiplus \Ppiminus}}
\newcommand{\eekkbkg}{\HepProcess{\Ppositron \Pelectron \to \Ppositron \Pelectron \PKplus \PKminus}}
\newcommand{\eeppbkg}{\HepProcess{\Ppositron \Pelectron \to \Ppositron \Pelectron \APproton \Pproton}}
\newcommand{\mumumumubkg}{\HepProcess{\Ppositron \Pelectron \to \APmuon \Pmuon \APmuon \Pmuon}}
\newcommand{\mumutautaubkg}{\HepProcess{\Ppositron \Pelectron \to \APmuon \Pmuon \APtauon \Ptauon}}
\newcommand{\tautautautaubkg}{\HepProcess{\Ppositron \Pelectron \to \APtauon \Ptauon \APtauon \Ptauon}}

\newcommand{\tautaubkg}{\HepProcess{\Ppositron \Pelectron \to \APtauon \Ptauon (\Pgamma)}}
\newcommand{\kkgammabkg}{\HepProcess{\Ppositron \Pelectron \to \PKplus \PKminus \Pgamma}}
\newcommand{\ksklgammabkg}{\HepProcess{\Ppositron \Pelectron \to \PKshort \PKlong \Pgamma}}
\newcommand{\pipigammabkg}{\HepProcess{\Ppositron \Pelectron \to \Ppiplus \Ppiminus \Pgamma}}
\newcommand{\pipipizgammabkg}{\HepProcess{\Ppositron \Pelectron \to \Ppiplus \Ppiminus \Ppizero \Pgamma}}
\newcommand{\hhgammabkg}{\HepProcess{\Ppositron \Pelectron \to hh \Pgamma}}

\newcommand{\eebkg}{\HepProcess{\Ppositron \Pelectron \to \Ppositron \Pelectron (\Pgamma)}}
\newcommand{\mumubkg}{\HepProcess{\Ppositron \Pelectron \to \APmuon \Pmuon (\Pgamma)}}
\newcommand{\ggbkg}{\HepProcess{\Ppositron \Pelectron \to \Pgamma \Pgamma (\Pgamma)}}
\newcommand{\bbbarbkg}{\HepProcess{\Ppositron \Pelectron \to \PB \APB}}
\newcommand{\continuumbkg}{\HepProcess{\Ppositron \Pelectron \to \Pquark \APquark}}

\newcommand{\aprimetochionechitwo}{\HepProcess{\aprime \to \chione \chitwo}}

\newcommand{\processmuon}{\HepProcess{\Ppositron \Pelectron \to \chione \chitwo (\to \chione \Ppositron \Pelectron) \hprime (\to \APmuon \Pmuon)}}
\newcommand{\processhadron}{\HepProcess{\Ppositron \Pelectron \to \chione \chitwo (\to \chione \Ppositron \Pelectron) \hprime}}

\newcommand{\hmumu}{\hprime{} \to \APmuon \Pmuon}
\newcommand{\hpipi}{\hprime{} \to \Ppiplus \Ppiminus}
\newcommand{\hkk}{\hprime{} \to \PKplus \PKminus}

\newcommand{\chitwoee}{\chitwo{} \to \Ppositron \Pelectron}
    
\newcommand{\mathdeltaalpha}{\Delta\alpha^{h^{\prime}}_{\vec{x}, \vec{p}}}
\newcommand{\mathlogdeltaalpha}{-\mathrm{log}(1 - \mathrm{cos}(\Delta\alpha^{h^{\prime}}_{\vec{x}, \vec{p}}))}
\newcommand{\deltaalpha}{$\mathdeltaalpha$}
\newcommand{\logdeltaalpha}{$\mathlogdeltaalpha$}

\newcommand{\mathemiss}{E_{\mathrm{miss}}}
\newcommand{\emiss}{$\mathemiss$}

\newcommand{\matheextra}{E_\mathrm{extra}(\mathrm{neutral})}
\newcommand{\eextra}{$\matheextra$}

\newcommand{\mathextratracks}{E_\mathrm{tracks}^\mathrm{ROE}}
\newcommand{\extratracks}{$\mathextratracks$}

\newcommand{\mathinvM}{M_\mathrm{inv}^{h^{\prime}}}
\newcommand{\invM}{$\mathinvM$}

\newcommand{\mathchitwoinvM}{M_\mathrm{ee}^{\chi_2}}
\newcommand{\chitwoinvM}{$\mathchitwoinvM$}

\newcommand{\maththetaKLM}{\theta^\mathrm{KLM,ext}}
\newcommand{\thetaKLM}{$\maththetaKLM$}

\newcommand{\mathmchione}{m_{\chione{}}}
\newcommand{\mchione}{$\mathmchione$}

\newcommand{\mathmchitwo}{m_{\chitwo{}}}
\newcommand{\mchitwo}{$\mathmchitwo$}

\newcommand{\mathmaprime}{m_{\aprime{}}}
\newcommand{\maprime}{$\mathmaprime$}

\newcommand{\mathmhprime}{m_{\hprime{}}}
\newcommand{\mhprime}{$\mathmhprime$}

\newcommand{\mathctauhprime}{c\tau_{\hprime{}}}
\newcommand{\ctauhprime}{$\mathctauhprime$}

\newcommand{\mathctauchitwo}{c\tau_{\chitwo{}}}
\newcommand{\ctauchitwo}{$\mathctauchitwo$}

\newcommand{\mathmixingepsilon}{\epsilon}
\newcommand{\mixingepsilon}{$\mathmixingepsilon$}

\newcommand{\mathmixingtheta}{\theta}
\newcommand{\mixingtheta}{$\mathmixingtheta$}

\newcommand{\mathf}{f}
\newcommand{\f}{$\mathf$}

\newcommand{\mathgx}{g_X}
\newcommand{\gx}{$\mathgx$}

\newcommand{\mathalphad}{\alpha_D}
\newcommand{\alphad}{$\mathalphad$}

\newcommand{\mathalphaf}{\alpha_f}
\newcommand{\alphaf}{$\mathalphaf$}

\newcommand{\mathdeltam}{\Delta m}
\newcommand{\deltam}{$\mathdeltam$}

\newcommand{\mathcdcextra}{N_\mathrm{extra}^\mathrm{CDC}}
\newcommand{\cdcextra}{$\mathcdcextra$}

\newcommand{\mathcrossx}{\sigma_\mathrm{sig}}
\newcommand{\crossx}{$\mathcrossx$}

\newcommand{\mathvertexfit}{\mathcal{P}_\mathrm{vertex}}
\newcommand{\vertexfit}{$\mathvertexfit$}

\newcommand{\mathdeltaalphaklm}{\Delta \alpha}
\newcommand{\deltaalphaklm}{$\mathdeltaalphaklm$}

\newcommand{\maththetamiss}{\theta_\mathrm{miss}^\mathrm{lab}}
\newcommand{\thetamiss}{$\maththetamiss$}


\definecolor{kit-green}{RGB}{0, 150, 130}
\definecolor{kit-green100}{RGB}{0, 150, 130}
\definecolor{kit-green90}{rgb}{0.1, 0.6294, 0.5588}
\definecolor{kit-green80}{rgb}{0.2, 0.6706, 0.6078}
\definecolor{kit-green75}{rgb}{0.25, 0.6912, 0.6324}
\definecolor{kit-green70}{rgb}{0.3, 0.7118, 0.6569}
\definecolor{kit-green60}{rgb}{0.4, 0.7529, 0.7059}
\definecolor{kit-green50}{rgb}{0.5, 0.7941, 0.7549}
\definecolor{kit-green40}{rgb}{0.6, 0.8353, 0.8039}
\definecolor{kit-green30}{rgb}{0.7, 0.8765, 0.8529}
\definecolor{kit-green25}{rgb}{0.75, 0.8971, 0.8775}
\definecolor{kit-green20}{rgb}{0.8, 0.9176, 0.902}
\definecolor{kit-green15}{rgb}{0.85, 0.9382, 0.9265}
\definecolor{kit-green10}{rgb}{0.9, 0.9588, 0.951}
\definecolor{kit-green5}{rgb}{0.95, 0.9794, 0.9755}

\definecolor{kit-blue}{RGB}{70, 100, 170}
\definecolor{kit-blue100}{RGB}{70, 100, 170}
\definecolor{kit-blue90}{rgb}{0.3471, 0.4529, 0.7}
\definecolor{kit-blue80}{rgb}{0.4196, 0.5137, 0.7333}
\definecolor{kit-blue75}{rgb}{0.4559, 0.5441, 0.75}
\definecolor{kit-blue70}{rgb}{0.4922, 0.5745, 0.7667}
\definecolor{kit-blue60}{rgb}{0.5647, 0.6353, 0.8}
\definecolor{kit-blue50}{rgb}{0.6373, 0.6961, 0.8333}
\definecolor{kit-blue40}{rgb}{0.7098, 0.7569, 0.8667}
\definecolor{kit-blue30}{rgb}{0.7824, 0.8176, 0.9}
\definecolor{kit-blue25}{rgb}{0.8186, 0.848, 0.9167}
\definecolor{kit-blue20}{rgb}{0.8549, 0.8784, 0.9333}
\definecolor{kit-blue15}{rgb}{0.8912, 0.9088, 0.95}
\definecolor{kit-blue10}{rgb}{0.9275, 0.9392, 0.9667}
\definecolor{kit-blue5}{rgb}{0.9637, 0.9696, 0.9833}

\definecolor{kit-red}{RGB}{162, 34, 35}
\definecolor{kit-red100}{RGB}{162, 34, 35}
\definecolor{kit-red90}{rgb}{0.6718, 0.22, 0.2235}
\definecolor{kit-red80}{rgb}{0.7082, 0.3067, 0.3098}
\definecolor{kit-red75}{rgb}{0.7265, 0.35, 0.3529}
\definecolor{kit-red70}{rgb}{0.7447, 0.3933, 0.3961}
\definecolor{kit-red60}{rgb}{0.7812, 0.48, 0.4824}
\definecolor{kit-red50}{rgb}{0.8176, 0.5667, 0.5686}
\definecolor{kit-red40}{rgb}{0.8541, 0.6533, 0.6549}
\definecolor{kit-red30}{rgb}{0.8906, 0.74, 0.7412}
\definecolor{kit-red25}{rgb}{0.9088, 0.7833, 0.7843}
\definecolor{kit-red20}{rgb}{0.9271, 0.8267, 0.8275}
\definecolor{kit-red15}{rgb}{0.9453, 0.87, 0.8706}
\definecolor{kit-red10}{rgb}{0.9635, 0.9133, 0.9137}
\definecolor{kit-red5}{rgb}{0.9818, 0.9567, 0.9569}

\definecolor{kit-yellow}{RGB}{252, 229, 0}
\definecolor{kit-yellow100}{RGB}{252, 229, 0}
\definecolor{kit-yellow90}{rgb}{0.9894, 0.9082, 0.1}
\definecolor{kit-yellow80}{rgb}{0.9906, 0.9184, 0.2}
\definecolor{kit-yellow75}{rgb}{0.9912, 0.9235, 0.25}
\definecolor{kit-yellow70}{rgb}{0.9918, 0.9286, 0.3}
\definecolor{kit-yellow60}{rgb}{0.9929, 0.9388, 0.4}
\definecolor{kit-yellow50}{rgb}{0.9941, 0.949, 0.5}
\definecolor{kit-yellow40}{rgb}{0.9953, 0.9592, 0.6}
\definecolor{kit-yellow30}{rgb}{0.9965, 0.9694, 0.7}
\definecolor{kit-yellow25}{rgb}{0.9971, 0.9745, 0.75}
\definecolor{kit-yellow20}{rgb}{0.9976, 0.9796, 0.8}
\definecolor{kit-yellow15}{rgb}{0.9982, 0.9847, 0.85}
\definecolor{kit-yellow10}{rgb}{0.9988, 0.9898, 0.9}
\definecolor{kit-yellow5}{rgb}{0.9994, 0.9949, 0.95}

\definecolor{kit-orange}{RGB}{223, 155, 27}
\definecolor{kit-orange100}{RGB}{223, 155, 27}
\definecolor{kit-orange90}{rgb}{0.8871, 0.6471, 0.1953}
\definecolor{kit-orange80}{rgb}{0.8996, 0.6863, 0.2847}
\definecolor{kit-orange75}{rgb}{0.9059, 0.7059, 0.3294}
\definecolor{kit-orange70}{rgb}{0.9122, 0.7255, 0.3741}
\definecolor{kit-orange60}{rgb}{0.9247, 0.7647, 0.4635}
\definecolor{kit-orange50}{rgb}{0.9373, 0.8039, 0.5529}
\definecolor{kit-orange40}{rgb}{0.9498, 0.8431, 0.6424}
\definecolor{kit-orange30}{rgb}{0.9624, 0.8824, 0.7318}
\definecolor{kit-orange25}{rgb}{0.9686, 0.902, 0.7765}
\definecolor{kit-orange20}{rgb}{0.9749, 0.9216, 0.8212}
\definecolor{kit-orange15}{rgb}{0.9812, 0.9412, 0.8659}
\definecolor{kit-orange10}{rgb}{0.9875, 0.9608, 0.9106}
\definecolor{kit-orange5}{rgb}{0.9937, 0.9804, 0.9553}

\definecolor{kit-lightgreen}{RGB}{140, 182, 60}
\definecolor{kit-lightgreen100}{RGB}{140, 182, 60}
\definecolor{kit-lightgreen90}{rgb}{0.5941, 0.7424, 0.3118}
\definecolor{kit-lightgreen80}{rgb}{0.6392, 0.771, 0.3882}
\definecolor{kit-lightgreen75}{rgb}{0.6618, 0.7853, 0.4265}
\definecolor{kit-lightgreen70}{rgb}{0.6843, 0.7996, 0.4647}
\definecolor{kit-lightgreen60}{rgb}{0.7294, 0.8282, 0.5412}
\definecolor{kit-lightgreen50}{rgb}{0.7745, 0.8569, 0.6176}
\definecolor{kit-lightgreen40}{rgb}{0.8196, 0.8855, 0.6941}
\definecolor{kit-lightgreen30}{rgb}{0.8647, 0.9141, 0.7706}
\definecolor{kit-lightgreen25}{rgb}{0.8873, 0.9284, 0.8088}
\definecolor{kit-lightgreen20}{rgb}{0.9098, 0.9427, 0.8471}
\definecolor{kit-lightgreen15}{rgb}{0.9324, 0.9571, 0.8853}
\definecolor{kit-lightgreen10}{rgb}{0.9549, 0.9714, 0.9235}
\definecolor{kit-lightgreen5}{rgb}{0.9775, 0.9857, 0.9618}

\definecolor{kit-purple}{RGB}{163, 16, 124}
\definecolor{kit-purple100}{RGB}{163, 16, 124}
\definecolor{kit-purple90}{rgb}{0.6753, 0.1565, 0.5376}
\definecolor{kit-purple80}{rgb}{0.7114, 0.2502, 0.589}
\definecolor{kit-purple75}{rgb}{0.7294, 0.2971, 0.6147}
\definecolor{kit-purple70}{rgb}{0.7475, 0.3439, 0.6404}
\definecolor{kit-purple60}{rgb}{0.7835, 0.4376, 0.6918}
\definecolor{kit-purple50}{rgb}{0.8196, 0.5314, 0.7431}
\definecolor{kit-purple40}{rgb}{0.8557, 0.6251, 0.7945}
\definecolor{kit-purple30}{rgb}{0.8918, 0.7188, 0.8459}
\definecolor{kit-purple25}{rgb}{0.9098, 0.7657, 0.8716}
\definecolor{kit-purple20}{rgb}{0.9278, 0.8125, 0.8973}
\definecolor{kit-purple15}{rgb}{0.9459, 0.8594, 0.9229}
\definecolor{kit-purple10}{rgb}{0.9639, 0.9063, 0.9486}
\definecolor{kit-purple5}{rgb}{0.982, 0.9531, 0.9743}

\definecolor{kit-brown}{RGB}{167, 130, 46}
\definecolor{kit-brown100}{RGB}{167, 130, 46}
\definecolor{kit-brown90}{rgb}{0.6894, 0.5588, 0.2624}
\definecolor{kit-brown80}{rgb}{0.7239, 0.6078, 0.3443}
\definecolor{kit-brown75}{rgb}{0.7412, 0.6324, 0.3853}
\definecolor{kit-brown70}{rgb}{0.7584, 0.6569, 0.4263}
\definecolor{kit-brown60}{rgb}{0.7929, 0.7059, 0.5082}
\definecolor{kit-brown50}{rgb}{0.8275, 0.7549, 0.5902}
\definecolor{kit-brown40}{rgb}{0.862, 0.8039, 0.6722}
\definecolor{kit-brown30}{rgb}{0.8965, 0.8529, 0.7541}
\definecolor{kit-brown25}{rgb}{0.9137, 0.8775, 0.7951}
\definecolor{kit-brown20}{rgb}{0.931, 0.902, 0.8361}
\definecolor{kit-brown15}{rgb}{0.9482, 0.9265, 0.8771}
\definecolor{kit-brown10}{rgb}{0.9655, 0.951, 0.918}
\definecolor{kit-brown5}{rgb}{0.9827, 0.9755, 0.959}

\definecolor{kit-cyan}{RGB}{35, 161, 224}
\definecolor{kit-cyan100}{RGB}{35, 161, 224}
\definecolor{kit-cyan90}{rgb}{0.2235, 0.6682, 0.8906}
\definecolor{kit-cyan80}{rgb}{0.3098, 0.7051, 0.9027}
\definecolor{kit-cyan75}{rgb}{0.3529, 0.7235, 0.9088}
\definecolor{kit-cyan70}{rgb}{0.3961, 0.742, 0.9149}
\definecolor{kit-cyan60}{rgb}{0.4824, 0.7788, 0.9271}
\definecolor{kit-cyan50}{rgb}{0.5686, 0.8157, 0.9392}
\definecolor{kit-cyan40}{rgb}{0.6549, 0.8525, 0.9514}
\definecolor{kit-cyan30}{rgb}{0.7412, 0.8894, 0.9635}
\definecolor{kit-cyan25}{rgb}{0.7843, 0.9078, 0.9696}
\definecolor{kit-cyan20}{rgb}{0.8275, 0.9263, 0.9757}
\definecolor{kit-cyan15}{rgb}{0.8706, 0.9447, 0.9818}
\definecolor{kit-cyan10}{rgb}{0.9137, 0.9631, 0.9878}
\definecolor{kit-cyan5}{rgb}{0.9569, 0.9816, 0.9939}

\definecolor{kit-gray}{RGB}{0, 0, 0}
\definecolor{kit-gray100}{RGB}{0, 0, 0}
\definecolor{kit-gray90}{rgb}{0.1, 0.1, 0.1}
\definecolor{kit-gray80}{rgb}{0.2, 0.2, 0.2}
\definecolor{kit-gray75}{rgb}{0.25, 0.25, 0.25}
\definecolor{kit-gray70}{rgb}{0.3, 0.3, 0.3}
\definecolor{kit-gray60}{rgb}{0.4, 0.4, 0.4}
\definecolor{kit-gray50}{rgb}{0.5, 0.5, 0.5}
\definecolor{kit-gray40}{rgb}{0.6, 0.6, 0.6}
\definecolor{kit-gray30}{rgb}{0.7, 0.7, 0.7}
\definecolor{kit-gray25}{rgb}{0.75, 0.75, 0.75}
\definecolor{kit-gray20}{rgb}{0.8, 0.8, 0.8}
\definecolor{kit-gray15}{rgb}{0.85, 0.85, 0.85}
\definecolor{kit-gray10}{rgb}{0.9, 0.9, 0.9}
\definecolor{kit-gray5}{rgb}{0.95, 0.95, 0.95}
    \newacronym{kek}{KEK}{High Energy Accelerator Research Organisation}
\newacronym{her}{HER}{high energy ring}
\newacronym{ler}{LER}{low energy ring}
\newacronym{hlt}{HLT}{High Level Trigger}
\newacronym{cmb}{CMB}{Comsmic Microwave Background}
\newacronym{depfet}{DEPFET}{Depleted Field Effect Transistor}
\newacronym{nn}{NN}{Neural Network}
\newacronym{ml}{ML}{Machine Learning}
\newacronym{cms}{CMS}{Center of Mass}

\newacronym{hep}{HEP}{High Energy Physics}
\newacronym{sm}{SM}{Standard Model}
\newacronym{dm}{DM}{Dark Matter}
\newacronym{bsm}{BSM}{physics beyond the Standard Model}
\newacronym{np}{NP}{New Physics}
\newacronym{qcd}{QCD}{quantum chromodynamics}
\newacronym{ip}{IP}{Interaction Point}
\newacronym{fsp}{FSP}{Final State Particle}
\newacronym{pid}{PID}{Particle Identification}
\newacronym{bcs}{BCS}{Best Candidate Selection}

\newacronym{isr}{ISR}{Initial State Radiation}
\newacronym{fsr}{FSR}{Final State Radiation}

\newacronym{pxd}{PXD}{Pixel Detector}
\newacronym{svd}{SVD}{Silicon Vertex Detector}
\newacronym{vxd}{VXD}{Vertex Detectors}
\newacronym{cdc}{CDC}{Central Drift Chamber}
\newacronym{top}{TOP}{Time-of-Propagation}
\newacronym{arich}{ARICH}{Aerogel Ring-imaging Cherenkov}
\newacronym{ecl}{ECL}{Electromagnetic Calorimeter}
\newacronym{klm}{KLM}{$K_L$ and Muon Detector}

\newacronym{mc}{MC}{Monte Carlo}
\newacronym{pdf}{PDF}{Probability Density Functions}

\newacronym{basf2}{\basftwoshort}{Belle~II Analysis Software Framework}

\newacronym{lm}{LM}{Local Maximum}
\newacronym{cr}{CR}{Connected Regions}
\newacronym{hdlm}{hdLM}{highest deposition Local Maximum}
\newacronym{hdp}{hDP}{highest deposition MCParticle}
\newacronym{mip}{MIP}{minimal ionizing particle}
\newacronym{psd}{PSD}{Puls Shape Discrimination}
\newacronym{pm}{PM}{Particle Matching}
\newacronym{mcm}{McM}{MCMatching}

\newacronym{gnn}{GNN}{Graph Neural Network}
\newacronym{swa}{SWA}{Stochastic Weighted Averaging}
\newacronym{mse}{MSE}{Mean Squared Error}

    \title{Using Graph Neural Networks for hadronic clustering and to reduce
    beam background in the Belle~II electromagnetic calorimeter}
    \maketitle
    \author{Jonas Eppelt$^{1}$,Torben Ferber$^{1}$}

    \affil{$^{1}$Institute of Experimental Particle Physics, Karlsruhe Institute of Technology, 76131 Karlsruhe, Germany}

    \email{jonas.eppelt@kit.edu}

    \begin{abstract}
        The Belle~II electromagnetic calorimeter consists of 8376 CsI(Tl)
        scintillation crystals and is not only used for measuring
        electromagnetic particles but also for identifying and determining the
        position of hadrons, particularly neutral\textbf{} hadrons. Recent data-taking
        periods have presented challenges for the current clustering method: Firstly,
        the record-breaking luminosities achieved by the SuperKEKB accelerator
        have increased background rates, leading to a higher number of crystals with
        energy depositions, and an overall increase in the total energy measured
        in the calorimeter. This resulted in poorer photon energy resolution and
        the reconstruction of more fake photon clusters. Secondly, challenges arise
        from the nature of hadronic interactions. In contrast to $\gamma$ and
        $e^{\pm}$, hadrons interacting in the calorimeter result in irregular,
        sometimes even disconnected energy depositions. These clusters can be misinterpreted
        as photon clusters, thereby reducing the position resolution of neutral
        hadrons or causing a complete misidentification of the hadron. Graph neural
        networks offer a promising solution to both challenges. By representing
        only crystals with an energy measurement as nodes, graphs capture the sparsity
        of the input. Using message-passing layers that learn the graph edges
        also helps to address the asymmetric sensor layout of Belle~II's ECL. In
        these proceedings, we will present a novel approach to identify the
        challenges in the detector simulation. Using this information, we train
        a Graph Neural Network to identify and remove unwanted depositions
        before clustering.
    \end{abstract}
    \section{Introduction}
    \label{ch:intro}

    The Belle~II experiment \cite{abeBelleIITechnical2010} is a multi-purpose
    particle detector located at the SuperKEKB accelerator in Tsukuba, Japan. It
    reached record-breaking instantaneous luminosities close to
    $5\times 10^{34}$ \SI{}{\per\centi\meter\squared\per\second}. With the accelerator
    colliding beams mostly on the \FourS resonance, the experiment mainly
    studies the decays of $B$-mesons. The detector consists of a three-part tracking
    system (\gls*{pxd}, \gls*{svd}, and \gls*{cdc}), an \gls*{ecl} for energy
    measurements, and particle identification systems (\gls*{top}, \gls*{arich},
    and the \gls*{klm}). The \gls*{ecl} consists of 8736 CsI(Tl) scintillation crystals,
    oriented in such a way that their faces point towards the interaction point
    with a slight tilt to prevent particles escaping through structural gaps
    between the crystal. The calorimeter is split into three sections: the
    forward endcap, the barrel region, and the backward endcap. The average crystal
    has a $6\times6$ \SI{}{\centi\meter\squared} face area and is about \SI{30}{\centi\meter}
    long. The light yield of each crystal is read out by two photodiodes, each connected
    to a preamplifier and is read out every \SI{0.56594}{\micro\second}. 31 read
    out points are used to describe the light pulse shape. Since hadronic showers
    produce a faster light pulse compared to electromagnetic particles, a multi-template
    fit is used to determine the amplitude and hadronic fraction of the pulse
    \cite{Longo:2020zqt}, as well as the pulse time.

    As many particles deposit energy in more than one crystal, the following
    clustering algorithm is currently employed to group those energy depositions
    \cite{haideRealTimeGraphNeural2025}. Starting from seed crystals ($>$ \SI{10}{\mega\eV}),
    \gls*{cr} are constructed by adding any directly neighboring crystal with
    more than \SI{10}{\mega\eV} and next-to-neighboring crystals with \SI{0.5}{\mega\eV}.
    If any \gls*{cr} share a crystal, they are merged into one. Within each \gls*{cr}
    \gls*{lm} are identified, which must have $>$ \SI{10}{\mega\eV} energy
    measured and more then any neighboring crystal. If a \gls*{cr} contains multiple
    \gls*{lm}, clusters are created by splitting the \gls*{cr} into as many
    clusters as \gls*{lm} are contained in it. For this, each crystal is
    assigned a fractional weight based on the energy and distance to the
    respective \gls*{lm}. The cluster centroid is updated iteratively, until it
    is stable within \SI{1}{mm}. Barring only a few exceptions,\footnote{If the
    centroid moves away too far from the original \gls*{lm} position, the cluster
    is discarded.} every \gls*{lm} leads to a cluster.

    The central role that \gls*{lm}s play in this algorithm makes it very
    susceptible to an increased number of them, which subsequently leads to several
    challenges:
    \begin{itemize}
        \item A larger number of \gls*{lm} increases runtimes of the clustering and
            subsequent algorithms.

        \item More \glspl*{lm} lead to more clusters to be stored, which
            increases file sizes.

        \item Clusters might be split accidentally when background processes create
            a second \gls{lm} in a \gls*{cr}.

        \item Many analyses in Belle~II make use of the fact that the collision energy
            is well known. Therefore, only as much energy as released in the collision
            should be found in the calorimeter. Additional clusters make this
            more difficult.
    \end{itemize}

    Two main processes lead to unnecessarily created \gls*{lm}. In pursuit of
    its target instantaneous luminosity of $\approx 5\times 10^{35}$\SI{}{\per\centi\meter\squared\per\second},
    the SuperKEKB accelerator has also increased so-called beam background
    processes \cite{natochiiBeamBackgroundExpectations2022}. These are background
    processes that correlate with the operation of the electron/positron beams,
    mainly:
    \begin{itemize}
        \item Particle scattering with residual gas molecules

        \item Inter-bunch scattering (Touschek effect)

        \item Radiative Bhabha scattering and two-photon processes

        \item Synchrotron radiation.
    \end{itemize}

    The second source of additional \gls*{lm}s are hadronic interactions in the \gls*{ecl}.
    These are less regularly shaped compared to pure electromagnetic
    interactions and can create so-called split-offs: particles (mostly neutrons)
    that are created during the shower and traverse a significant distance before
    interacting again. This creates separate clusters that the current
    clustering algorithm cannot attribute to the original process.

    \section{Local Maximum Classification}
    To counter the increasing number of \glspl*{lm}, we propose a \gls*{gnn} based
    classification algorithm that identifies \glspl*{lm} originating from background
    processes before \gls*{cr} and clusters are created.

    \subsection{Training target labels}
    To properly evaluate any classification algorithm, we first need to identify
    the various types of \glspl*{lm}. For this, we use information provided by
    \gls{basf2} \cite{thebelleiicollaborationBelleIIAnalysis2025} during the detector
    simulation. The simulation uses three components. The event generator first provides
    particles with kinematic properties resulting from the collision process.
    \gfour
    \cite{agostinelliGeant4aSimulationToolkit2003, allisonGeant4DevelopmentsApplications2006,
    allisonRecentDevelopmentsGeant42016}
    then simulates both the material interactions and the detector response of
    Belle II. Beam background is incorporated by overlaying data from randomly triggered
    events, which sample the detector response in the absence of physics signals,
    onto simulated events . Any \gls*{lm}, that has less than \SI{20}{\mega\eV}
    deposited in it by simulated particles is labeled as \textit{beam background}.
    Subsequently, each simulated particle is matched to the \gls*{lm} in which it
    deposited the highest energy. This ensures that there is one main \gls*{lm}
    per particle in cases where multiple \glspl*{lm} were created by it, e.g.,
    hadronic interactions. Any \gls*{lm} not associated with a simulated particle
    is labeled as a \textit{duplicate \gls*{lm}}. Lastly, we label all \glspl*{lm}
    associated with a split-off particle as \textit{split-off} and the remaining
    \glspl*{lm} as \textit{signal}. The final classification task is to identify
    \textit{signal} \glspl*{lm} against all other kinds of background \glspl*{lm}:
    \textit{beam background}, \textit{wrongly split clusters}, and \textit{split-off}.

    \subsubsection{Definition of split-off particles}
    Any particle simulated falls in one of two broad categories. Primary particles
    are those created by event generators and typically include final-state particles
    like photons, muons, or kaons. Particles that are created during the simulation
    of the detector response and material interactions (using \gfour) are called
    secondaries. This includes particles created during the electromagnetic
    shower in the \gls*{ecl} or by other material interactions like pair
    conversions. After the simulation, however, most secondary particles are not stored
    individually. Instead, their energy depositions are attributed to the primary
    particle they descended from. Only secondaries fulfilling one of the
    following conditions are stored:
    \begin{enumerate}

        \item It is the result of an in-flight decay (e.g. $K_{S}^{0}\rightarrow
            \pi^{+}\pi^{-}$).

        \item It creates \textit{relevant} hits in any subdetector system; what qualifies
            as a \textit{relevant} hit is decided on a per-subdetector basis, and
            as of now, the following conditions are implemented:
            \begin{itemize}
                \item The secondary has at least a simulated track of $\SI{15}{\centi\meter}$
                    and at least one hit in the \gls*{cdc}.

                \item The secondary has at least one hit in the \gls*{svd} or \gls*{pxd},
                    and the simulated track left the sub-detector area.
            \end{itemize}
    \end{enumerate}

    None of these conditions applies to split-off particles created by hadronic
    interactions in the \gls*{ecl} and all of their energy depositions are
    attributed to the original particle. Therefore, we introduce two new
    conditions to store secondaries:
    \begin{itemize}
        \item Secondaries that are created in the \textit{inner} part of Belle~II
            and that have a kinetic energy
            $E_{\text{kin}}> E_{\text{kin}}^{thr}$.

        \item Secondaries that are created in the \text{outer} part of Belle~II and
            have a kinetic energy $E_{\text{kin}}> E_{\text{kin}}^{thr}$ and have
            a distance between their production and decay vertex (given by
            \gfour) of over \SI{40}{\centi\meter}.
    \end{itemize}
    To distinguish the \textit{inner} and \textit{outer} part of the detector,
    we define these geometric boundaries:
    \begin{itemize}
        \item The distance to the z-axis $\rho_{\text{border}}= \SI{114}{\centi\meter}$.

        \item The distance from the interaction in the forward direction $z_{\text{forward}}
            = \SI{184}{\centi\meter}$.

        \item The distance from the interaction point in the backward direction $z
            _{\text{backward}}= \SI{-92}{\centi\meter}$.
    \end{itemize}
    Since we observed numerous showers starting before the \gls*{ecl}, we base
    the border values on the \gls*{top} detector located just before it. The more
    detailed information from the enables me to do a more comprehensive classification
    of simulated particles. Firstly, we define \textit{findable} particles as
    particles that have at least \SI{20}{\mega\eV} energy deposited in a \gls*{lm}.
    A \textit{root} particle is any primary particle as well as the daughters of
    particles commonly decaying in the Belle~II detector volume ($\pi^{0}$,
    $K_{S}^{0}$, $\Lambda$, $\eta$). To capture cases where \textit{root} particles
    are unfindable due to technical reasons, we define \textit{proxy} particles.
    These are descendants of an unfindable \textit{root} particle that are findable
    and whose momentum vector has a maximal angle of \SI{80}{\degree} to the momentum
    of the \textit{root} particle. If multiple descendants fulfill these criteria,
    the one with the smallest angle is the \textit{proxy}. All remaining
    particles are called \textit{split-off} particles.

    \subsection{Model architecture and training}
    %
    The network's architecture follows closely existing, related work
    \cite{ haideRealTimeGraphNeural2025, Belle-II:2023cal, Qasim:2019otl}: A global
    exchange appends graph-wide averages for each input feature to the node features.
    Then, a BatchNormalization layer and two linear layers calculate node embeddings.
    A stack of message passing blocks with skip connections feeds into a global
    mean pool to build graph embeddings. Here, each message passing block
    consists of a feed-forward layer, a modified GravNet \cite{juGraphNeuralNetworks2020}
    without the K-nearest-neighbor calculation, and a BatchNormLayer. The modified
    GravNetLayers project the nodes into a latent space and use the distance between
    them as a weight in the message passing. Skipping the K-nearest-neighbor
    calculation greatly improves training speeds and improves convergence and
    training stability. However, this requires the input graph to have pre-calculated
    edges. As the graphs we input are rather small, we use unweighted edges
    between all nodes. Finally, several feed-forward layers provide the
    classification output.

    We train the network using 20,000 $\FourS \rightarrow B \bar{B}$ events
    simulated using \textit{EvtGen} \cite{langeEvtGenParticleDecay2001} and 10,000
    events of $e^{+}e^{-}\rightarrow \phi(\rightarrow K_{L}^{0}K_{S}^{0}) \gamma$
    simulated using \phokhara \cite{campanarioStandardModelRadiative2019}. For
    each \gls*{lm}, the surrounding neighbors up to the 5th order (corresponding
    to a 9x9 square around the \gls*{lm} in the barrel) are used to form the input
    graph. Nodes in this graph represent crystals with th following features: the
    measured energy, time and detailed information from the pulse-shape template
    fit. Lastly, we use the crystal's mass to represent the different crystal
    geometries. Since our input graphs are relatively small, we construct them
    as fully connected.

    Further, we employ \gls*{swa} \cite{izmailovAveragingWeightsLeads2019} after
    100 epochs to further stabilize the training. As a loss function, we use the
    mean squared error as it shows a more stable behaviour than binary cross-entropy
    in our case.

    \subsection{Threshold determination}
\begin{figure}
        \centering
        \includegraphics[width=0.8\textwidth]{
            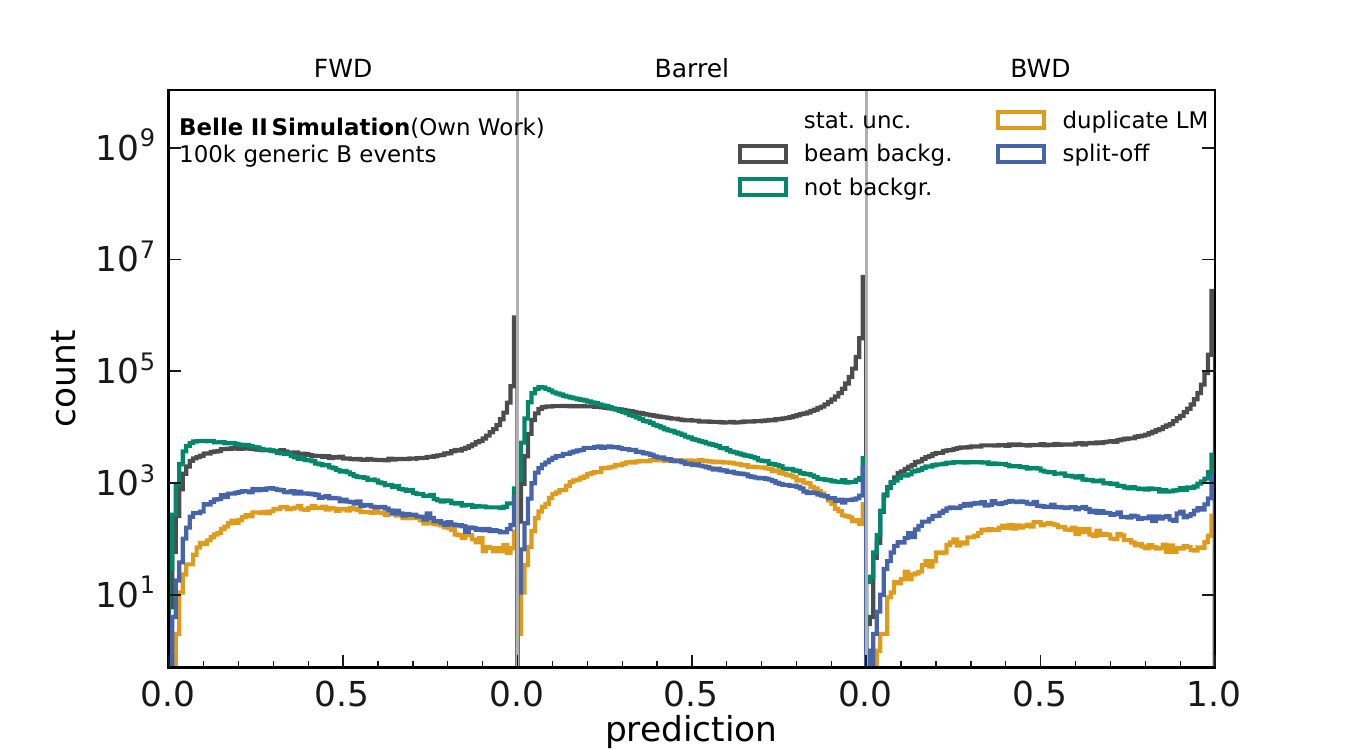
        }
        \caption{ This plot shows the overall classifier output distribution for
        the full energy range of \gls*{lm} in $\FourS \rightarrow B \bar{B}$ events.
        We show the distributions separately for the three detector regions: forward
        ("FWD"), barrel ("Barrel"), and backward (BWD). Further, we show the three
        true background classes beam background (beam backg.), \textit{split off}
        (split-off), and \textit{wrongly split clusters} ( wrongly sp.)
        separately. }
        \label{fig:output_distribution}
    \end{figure}

    As \cref{fig:output_distribution} shows, the majority of the beam background-induced
    \gls*{lm} are clearly identifiable, while the other background classes are more
    challenging as their distribution is less separated from the signal.
    Therefore, we decided to set the classifier working point to a fixed signal rejection
    rate of 5\%. To not be biased by the high number of \glspl*{lm} with low energy,
    we calculate the working point in bins of measured energy in the \gls*{lm}. As
    the background levels in the \gls{ecl} regions are very different, we also do
    the calculation separately for each detector region. Further, we apply the classifier
    only on \glspl*{lm} with less than \SI{100}{\mega\eV} and keep all above this,
    as the loss in physics performance for high energetic particles would not be
    acceptable. To smooth the cut values, we use cubic spline interpolation
    between bin centers. We determine the working points using an independent sample
    of $\FourS \rightarrow B \bar{B}$ events. The beam background overlay files
    are chosen to have the same conditions as the ones used for the training, and
    to be distinct from the ones used for the simulation of the training samples.

    \section{Results}
    To evaluate the classifier, we use simulated $\FourS \rightarrow B \bar{B}$
    events with beam background overlays taken under the same conditions used
    for training. Since the classifier is only applied for \glspl*{lm} with a measured
    energy between 20 and \SI{100}{\mega\eV}, we focus on evaluating in this
    energy range.
    \begin{figure}
        \centering
        \includegraphics[width=0.8\textwidth]{
            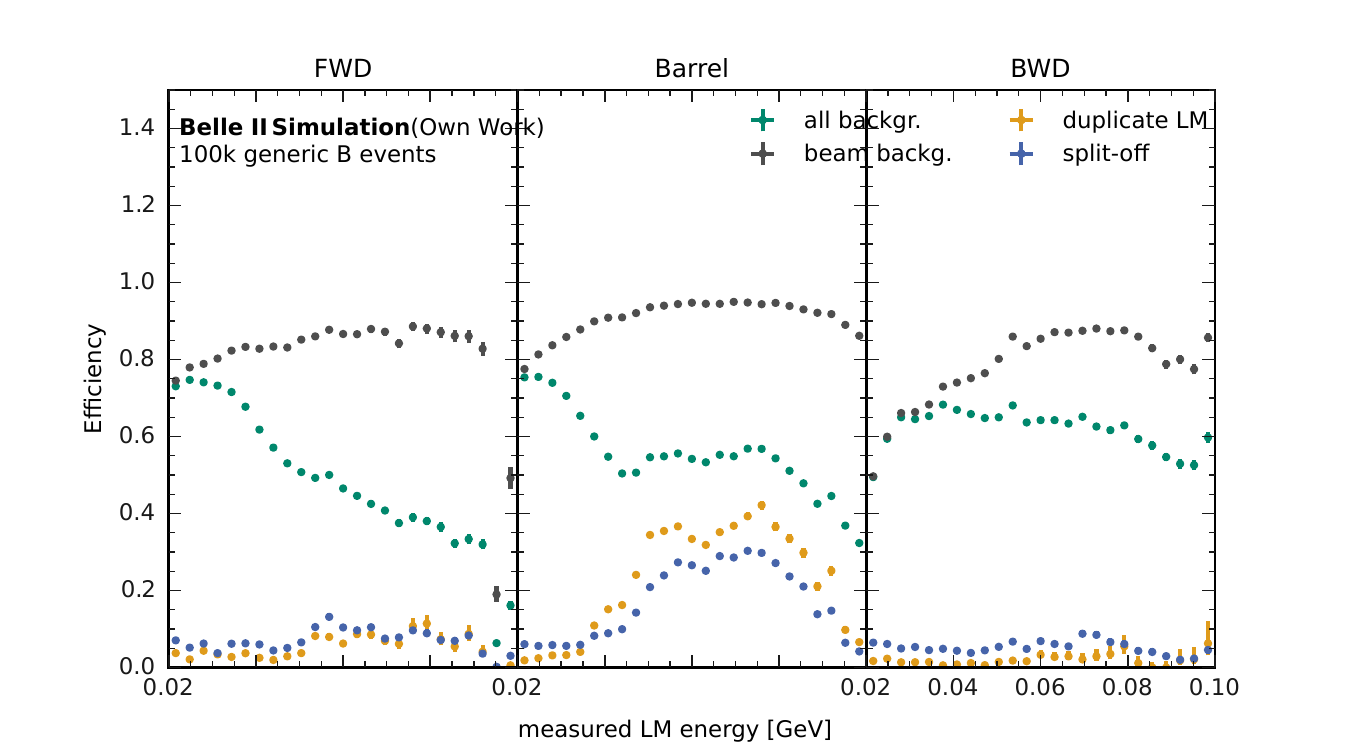
        }
        \caption{ Resulting rejection efficiencies of the classifier for each detector
        region in bins of the measured \gls*{lm} energy in generic $B$ events.
        We show them separately for each background class and the efficiency for
        all classes together. }
        \label{fig:results}
    \end{figure}
    As shown in \cref{fig:results}, we report that up to 95\% of beam background
    \glspl*{lm} can be removed by the classifier. In contrast, \textit{wrongly
    split clusters} and \textit{split-offs} prove to be more challenging. We
    achieve the highest rejection rate in the barrel above \SI{50}{\mega\eV}
    with about 40\%. This is because only above ca. \SI{50}{\mega\eV} the puls shape
    fit becomes precise enough to deliver reliable timing and pulse-shape information.
    In the endcap regions, especially the backward region, the efficiency for
    non-beam-background-induced \glspl*{lm} is significantly lower. The dominant
    reason for this is the high background level, making the separation from the
    true signal more challenging. In general, separating \textit{split-offs} and
    \textit{wrongly split clusters} from true signal \glspl*{lm} is challenging
    due to their very similar signatures and the fact that the underlying
    physical processes are basically identical: in all three cases, low-energetic
    particles, mostly neutrons or photons, create a shower in an \gls*{ecl}
    crystal.

    When averaging over all background types, our classifier reaches around 75\%
    rejection efficiency at low \gls*{lm} energies. This drops to around 30\% in
    the forward and barrel region, as the number of beam background induced \gls*{lm}
    decreases in these regions compared to the other background sources. In the
    backward region, the efficiency starts near 50\% and remains around 50-60\% throughout
    the relevant energy region. Here, the high number of beam background \gls*{lm}
    dominates over the other background sources.

    \section{Conclusions and outlook}
    We present an approach to reliably identify challenging signatures in Belle~II's
    \gls*{ecl} in simulated data. \textbf{}Namely, we can identify \glspl*{lm} created by
    beam background, secondary \gls*{lm} of a signal particle, and \gls*{lm} from
    (hadronic) \textit{\textit{split-offs}}. We use this information to label training
    samples and train a \gls*{gnn} to identify and remove background-induced \glspl*{lm}
    For purely beam background induced \glspl*{lm}, we reach a removal
    efficiency of around 90\% at a fixed signal efficiency of 95\%. For the other
    two classes, we reach up to 40\% in the barrel region between \SI{50}{\mega\eV}
    and \SI{100}{\mega\eV}. These results highlight the benefits of detailed
    studies of detector simulation when building training targets. We expect these
    studies to also benefit future endeavors in building \gls*{gnn} based cluster
    energy predictions.

    \printbibliography
\end{document}